\documentclass[twocolumn]{article}%aucune page-titre (pour ArXiv)
\usepackage{graphicx}
\usepackage{authblk}

\newcommand{\unit}[1]{\mathrm{~#1}}%unit�s/texte dans mode math
\newcommand{\moy}[1]{\left < #1 \right >}
\newcommand{\inp}[1]{\left(#1\right)}
\newcommand{\insb}[1]{\left[#1\right]}

\newcommand{\abs}[1]{\left|#1\right|}

\newcommand{\ioff}[1]{i\inp{#1}}
\newcommand{\icorr}[2]{\moy{\ioff{#1}\ioff{#2}}}

\newcommand{\intd}{\mathrm{d}}
\newcommand{\der}[3][]{\frac{\intd ^{#1}#2}{\intd #3^{#1}}}
\newcommand{\ee}[1]{\times 10^{#1}}
\newcommand{\vdc}{V_{\mathrm{dc}}}
\newcommand{\vac}{V_{\mathrm{ac}}}

\begin{document}

\title{Emission of Microwave Photon Pairs by a Tunnel Junction}

\author[1]{Jean-Charles Forgues\thanks{jean-charles.forgues@usherbrooke.ca}}
\author[1]{Christian Lupien\thanks{christian.lupien@usherbrooke.ca}}
\author[1]{Bertrand Reulet\thanks{bertrand.reulet@usherbrooke.ca}}
\affil[1]{D\'{e}partement de Physique, Universit\'{e} de Sherbrooke

Sherbrooke, Qu\'{e}bec, Canada, J1K 2R1}

\maketitle

\textbf{Generation and control of non-classical electromagnetic fields is of crucial importance for quantum information physics. While usual methods for the production of such fields rely on a non-linearity (of a crystal, a Josephson junction, etc.), a recent experiment performed on a normal conductor, a tunnel junction under microwave irradiation, has unveiled an alternative: the use of electron shot noise in a quantum conductor\cite{PAN_squeezing}. Here we show that such a device can emit \emph{pairs of microwave photons} of different frequencies with a rate as high as that of superconducting Josephson junctions\cite{Flurin}. This results in intensity fluctuations of the photon field at two different frequencies being correlated below the photon shot noise \emph{i.e.} two-mode amplitude squeezing. Our experiment constitutes a fundamental step towards the understanding of electronic noise in terms of quantum optics, and shows that even a normal conductor could be used as a resource for quantum information processing.}

While a classical current generates a classical field \cite{Glauber}, also known as a coherent state of light, it appears that when electron transport requires a quantum mechanical description, the random electromagnetic field that corresponds to current noise is non-classical. This was shown in \cite{PAN_squeezing} by the observation of vacuum noise squeezing. Squeezed electromagnetic fields are usually the result of non-linearities which appear in the Hamiltonian describing the electromagnetic field by terms such as $a^2$ and $a^{\dagger 2}$, with $a$ and $a^{\dagger}$ the photon annihilation and creation operators, \emph{i.e.} by emission/absorption of \emph{pairs} of photons \cite{Loudon}. Since a tunnel junction under appropriate dc and ac bias can emit a squeezed electromagnetic field despite its linearity, it is natural to consider whether the field it generates contains pairs of photons. It is precisely the goal of the present letter to address this question experimentally.

Another recent experiment has demonstrated that photo-assisted noise may exhibit correlations between the power fluctuations measured at two different frequencies for an adequate choice of the excitation frequency \cite{C4classique}. Since this experiment was performed at a relatively high temperature $T=3\unit{K}$ with regard to the frequency range, $4-8\unit{GHz}$, \emph{i.e.} $k_BT\gg hf$, the power fluctuations were classical, meaning that the measured correlations corresponded to fluctuations of photon fluxes with many ($\sim40$) photons emitted within an experimental detection window. Here we report the observation of similar correlations at very low temperature $T=20\unit{mK}$ and under weak excitation. We demonstrate that there still are correlations between power fluctuations measured at two frequencies, $f_1$ and $f_2$, even when in the quantum regime $k_BT\ll hf_{1,2}$ and when the average number of photons observed within a detection window is smaller than one. This is strong evidence for photons of frequencies $f_1$ and $f_2$ being emitted as a pair when the junction is irradiated at frequency $f_1+f_2$. Our data are in very good agreement with the theoretical predictions based on the fourth cumulant of current fluctuations in the quantum regime \cite{C4classique}.

\emph{Principle of the experiment}
\begin{figure}
    \includegraphics[width=\columnwidth] {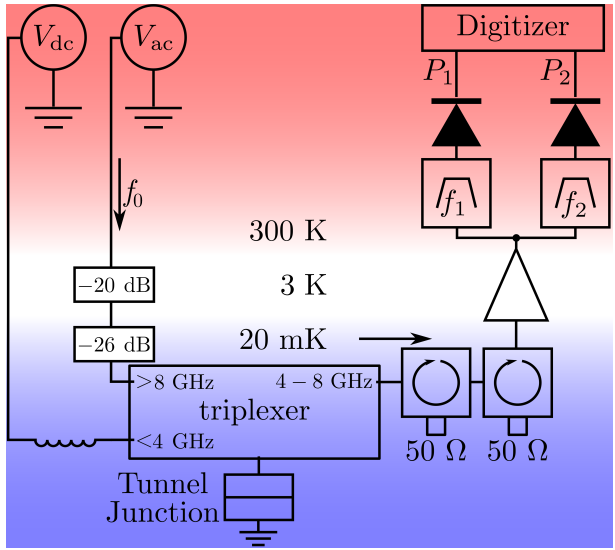}
    \caption{\footnotesize Experimental setup. The sample is a $23.6\unit{\Omega}$ Al/Al$_2$O$_3$/Al tunnel junction in the presence of a magnetic field to insure that the aluminum remains a normal metal at all temperatures. A triplexer seperates the frequency spectrum in three bands corresponding to the dc bias ($<4\unit{GHz}$), the ac bias ($>8\unit{GHz}$) and the detection ($4-8\unit{GHz}$). The noise generated by the junction in the $4-8\unit{GHz}$ range is amplified then separated in two frequency bands centered at $f_1=4.4\unit{GHz}$ and $f_2=7.2\unit{GHz}$ with bandwidths $\Delta f_1=0.65\unit{GHz}$ and $\Delta f_2=0.38\unit{GHz}$. The powers $P_{1,2}$ were measured with fast power detectors (diode symbols) with a $1\unit{ns}$ response time and digitized at a rate of $400\unit{MS/s}$ to compute $G_2=\moy{P_1P_2}-\moy{P_1}\moy{P_2}$ in real time.
    \label{figMontage}}
\end{figure}

The power $P(t)$ of the electromagnetic field radiated by a tunnel junction in a given frequency band centered on frequency $f$ fluctuates in time: $P(t)=\moy{P}+\delta P(t)$. The average power $\moy{P}$ is related to the spectral density $S(f)$ of the current fluctuations in the sample at frequency $f$ by $\moy{P}=G\inp{f}\insb{S\inp{f}+S_a\inp{f}}\Delta f$, where $G\inp{f}$ is the gain of the setup illustrated in Fig. \ref{figMontage}, $\Delta f$ the detection bandwidth and $S_a\inp{f}$ the noise spectral density of the HEMT amplifier placed at the $3\unit{K}$ stage. We measure the correlation $G_2=\moy{\delta P_1\delta P_2}$ between the power fluctuations $\delta P_{1,2}$ in two separate, non-overlapping frequency bands centered on $f_1$ and $f_2$ as a function of the dc and ac bias of the junction. The detection frequencies are chosen in the 4-8 GHz range, such that  $hf_{1,2}\gg k_BT$, \emph{i.e.} where quantum properties of the radiated field are prominent. Details on the calibration procedure are given in the Method section.

\emph{Results}

As in \cite{C4classique}, we observe $G_2\neq0$ only for excitation frequencies $f_0$ which respect $f_0=\inp{f_1\pm f_2}/p$ with $p$, an integer. In the following we will focus only on $f_0=f_1+f_2=11.6\unit{GHz}$, which best corresponds to the quantum regime.

\begin{figure}
  \includegraphics[width=\columnwidth] {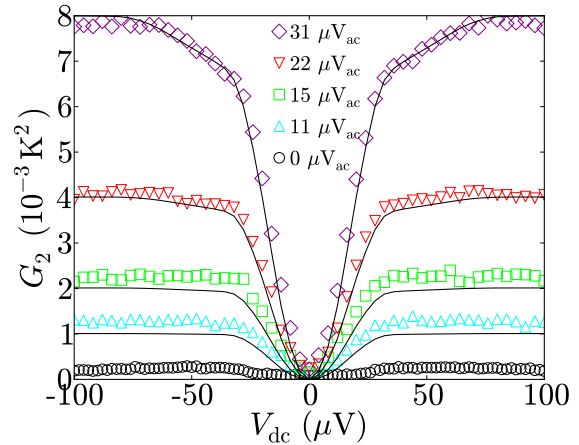}
  \caption{\footnotesize Reduced power-power correlator. Plot of $G_2$ vs dc bias voltage for various ac excitation amplitudes at frequency $f_0=f_1+f_2=11.6$ GHz. Symbols are experimental data and solid lines theoretical expectations of Eqs. (\ref{eqC4},\ref{eqX}). The overall gain was obtained by fitting the value of $G_2$ at large ac and dc bias (classical limit).
    \label{figFplus}}
\end{figure}

We show on Fig. \ref{figFplus} the result for $G_2$, expressed in units of $\mathrm{K}^2$, as a function of the dc bias voltage $V_\mathrm{dc}$ (coloured symbols), for various ac voltages $\vac$. The data exhibits kinks at integer multiples of $\vdc=hf_2/e\approx 30\unit{\mu V}$, which are characteristic of quantum effects related to emission or absorption of photons \cite{Lesovik}. Solid lines are the theoretical predictions for $G_2$ in terms of the fourth cumulant of current fluctuations $C_4$ given by \cite{C4classique}:

\begin{equation}
C_4=\abs{\icorr{f_1}{f_2}}^2
\label{eqC4}
\end{equation}

The two-frequency current-current correlator, which describes the noise dynamics under ac excitation, is given by \cite{GR1,GR2,GR3}:

\begin{equation}
\icorr{f}{f_0-f}=\sum_n \frac{\alpha_n}{2}[S_0\inp{f_{n+}}-S_0\inp{f_{n-}}]
\label{eqX}
\end{equation}
with $S_0\inp{f}=\inp{hf/R}\coth\inp{hf/2k_BT}$, the equilibrium noise spectral density at frequency $f$ in a tunnel junction of resistance $R$, $f_{n\pm}=f+nf_0\pm e\vdc/h$ and $\alpha_n=J_n\inp{e\vac/hf_0} J_{n+1}\inp{e\vac/hf_0}$, where $J_n$ are the Bessel functions of the first kind.

As evidenced by Fig. \ref{figFplus}, our data match very well with the theoretical expectation, thereby validating the proportionality between $G_2$ and $C_4$ in the quantum regime. The only difference between theory and experiment is that we observe a small extra contribution to $G_2$, which is most noticeable at very low ac bias. In particular, the photo-assisted $G_2$ should be zero in the absence of ac excitation while we measure a tiny contribution of magnitude $\sim2\ee{-4}\unit{K^2}$. This might be due to an imperfect calibration of the setup, but could also correspond to a real signal. Indeed, the current fluctuations generated by the junction are not Gaussian, which causes the existence of an intrinsic fourth cumulant, given by $e^3 V \Delta f/\inp{k_B^2G}\sim1.5\ee{-5}\unit{K^2}$ at $\vdc=100\unit{\mu V}$, \emph{i.e.} smaller than the observed signal by more than an order of magnitude.
As is the case for the third moment\cite{S3BR, Beenakker, Kindermann}, environmental effects can contribute to the fourth cumulant. In particular, a contribution $\sim\inp{\der{S}{V}}^2\moy{\delta V^2}$ is expected. Here, $\inp{\der{S}{V}}$ stands for the noise susceptibility \cite{GR1,GR2,GR3} and $\moy{\delta V^2}$ the voltage noise experienced by the sample within the relevant bandwidth. Considering low-frequency $\inp{\leq400\unit{MHz}}$ fluctuations of a $50\unit{\Omega}$ environment, a noise temperature of $5\unit{K}$ would be required to explain the observed signal. Further experimental study is required in order to fully explore this phenomenon.

\emph{Photon-Photon Correlations.}

The power detectors used in our experiment are not photo-detectors, but are sensitive to the total electric field generated by the sample. The detected power thus contains the contributions of photons emitted and absorbed by the junction as well as vacuum fluctuations. The amplifier noise adds a large contribution to this, which contributes to the average power but not to $G_2$. We will now evaluate $G_2$ in terms of photons emitted by the junction.

The total noise spectral density we detect can be decomposed into $S=S_{em}+Ghf$ where $S_{em}=2Ghf\moy{n(f)}$ is the emission noise \cite{LesovikLoosen,Gavish,Aguado} and $\moy{n(f)}$ is the average number of photons emitted at frequency $f$ per unit time per unit bandwidth. At equilibrium, one obtains the average number of photons emitted from $S_0\inp{f}$: $n_0\inp{f}=[\exp(hf/k_BT)-1]^{-1}$, here $\sim10^{-8}$, which corresponds to the Bose-Einstein distribution, as expected for thermal radiation.

\begin{figure}
  \includegraphics[width=\columnwidth] {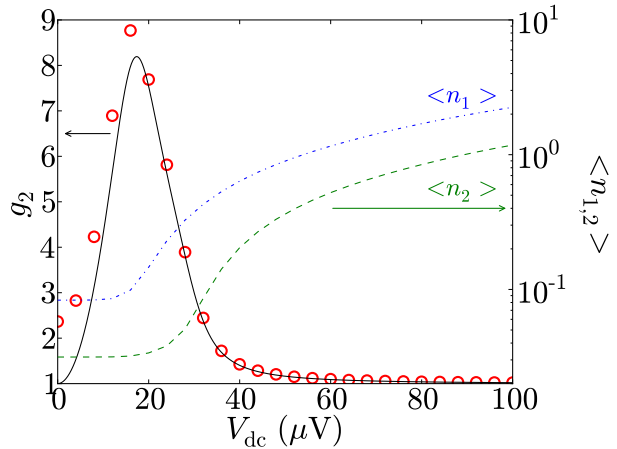}
  \caption{\footnotesize Photon-photon correlator. \emph{Left scale}: $g_2$ vs dc bias voltage at $V_{\mathrm{ac}}=22\unit{\mu V}$ for an excitation frequency. Red circles are experimental data and the full line represents theoretical expectations from Eq. (\ref{eqg2}). \emph{Right scale} (dashed lines): Number of emitted photons per unit bandwidth per unit time $\moy{n_{1,2}}$ in branches 1 and 2. They are calculated from the noise spectral density under ac excitation using $S\inp{f}=\frac{1}{2}\Sigma_n J_n^2(e\vac/hf_0)\insb{S_0\inp{f_{n+}}+S_0\inp{f_{n-}}}$. The number of photons detected in each measurement, $\moy{n}\Delta f \tau$, is close to $\moy{n}$ given that $\Delta f_1\tau=1.65$ and $\Delta f_2\tau=0.95$.
  \label{fig_g2}}
\end{figure}

Similarly, we can express the power-power correlator $G_2$ in terms of a photon-photon correlator. Using $\moy{\delta P_1\delta P_2}\propto\moy{\delta n_1\delta n_2}=\moy{n_1n_2}-\moy{n_1}\moy{n_2}$ with $n_{1,2}=n(f_{1,2})$, we define the normalized correlator\cite{Loudon,Iskhakov,Agafonov}:
\begin{equation}
g_2=\frac{\moy{n_1n_2}}{\moy{n_1}\moy{n_2}}=1+\frac{G_2}{S_{em}\inp{f_1}S_{em}\inp{f_2}}
\label{eqg2}
\end{equation}
which is shown by the red circles in Fig. \ref{fig_g2}, left scale.

While $g_2=1$ corresponds to photons at frequencies $f_1$ and $f_2$ being emitted independently, as for chaotic light, a value above 1 like the one observed here indicates the existence of correlations in the emission of photons \cite{Loudon}. When the number of emitted photons is large, correlations can be of classical origin, as in \cite{C4classique}. However, $g_2>1$ for $\moy{n_{1,2}}\ll1$, which can be seen in the large peak we observe on Fig. \ref{fig_g2}, implies the existence of correlations at the single photon level, \emph{i.e.} photon pairs. The average photon numbers $\moy{n_{1,2}}$ are plotted on the right scale of Fig. \ref{fig_g2}. Indeed, the peak in $g_2$ coincides with small photon numbers $\moy{n_1}\simeq 0.11$ and $\moy{n_2}\simeq 0.03$. If photons of frequency $f_2$ were always part of a pair, one would have $\moy{n_1n_2}=\moy{n_2}$ and $g_2=1/\moy{n_1}=9.1$. As we discuss below, this is almost the case.

In order to quantify the probability that a photon detected by our setup was emitted as part of a pair, we adopt a simple model valid in the $\moy{n_{1,2}}\ll1$ limit: we neglect the possibility of two photons reaching the same detector within a detection window. In that case, probabilities of detecting one photon at frequencies $f_1$ and $f_2$ within a detection window are respectively $P\inp{1}=\moy{n_1}$ and $P\inp{2}=\moy{n_2}$ while that of detecting a pair of photons is given by $P\inp{1,2}=\moy{n_1n_2}$. The probability of detecting a pair of photons when a photon is detected at $f_2$ is therefore $ P\inp{1|2}=\moy{n_1n_2}/\moy{n_2}$. The data in Fig. \ref{fig_g2} yields a probability of 94\% at $\vdc=16\unit{\mu V}$, $\vac=22\unit{\mu V}$ that a photon detected at $f_2$ was emitted as part of a pair. The creation of photon pairs of different frequencies in the microwave domain has been achieved recently with the help of superconducting circuits \cite{Flurin, Eichler, Nguyen}. Specifically, \cite{Flurin} reported a yield of $6\times10^6$ pairs per second, while our junction generates the same amount of pairs for a bandwidth of $200\unit{MHz}$. Since the two photons have different frequencies, a (purely dispersive) diplexer can be used to separate them spatially without loss.

\emph{Noise reduction factor}

\begin{figure}
  \includegraphics[width=\columnwidth] {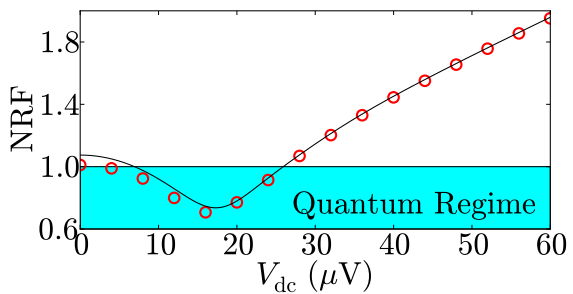}
  \caption{\footnotesize Noise reduction factor. Measured NRF vs dc bias voltage at $V_{\mathrm{ac}}=22\unit{\mu V}$ for an excitation frequency $f_0=f_1+f_2=11.6\unit{GHz}$ (red circles). The full line represents theoretical expectations.
  \label{fig_NRF}}
\end{figure}

The previous description corresponds to coincidence measurements in optics. Our observations can also be viewed as evidence for two-mode amplitude squeezing, \emph{i.e.} the ability for two light beams to have relative intensity fluctuations below the classical limit \cite{Iskhakov}. Classically, the variance of $n_1-n_2$ is limited by the photon shot noise of the two beams, given by $\moy{n_1}+\moy{n_2}$. Thus, one usually defines the noise reduction factor NRF as \cite{Heidmann,Debuisschert,Agafonov}:

\begin{equation}
NRF=\frac{\langle(\delta n_1-\delta n_2)^2\rangle}{\langle n_1\rangle + \langle n_2\rangle}
\end{equation}
which is greater than 1 for classical light. To calculate the NRF for our experiment, one needs to know the variance of the photon number fluctuations at each frequency $\moy{\delta n_{1,2}^2}$. 

This quantity is difficult to ascertain experimentally given the large contribution of the amplifier. However, the voltage noise measured in each frequency band is almost Gaussian, \emph{i.e.} identical to that of thermal noise of a macroscopic resistor, whose photon statistics is that of chaotic light. Thus it obeys $\moy{\delta n^2}=\moy{n}\inp{\moy{n}+1}$ \cite{BeenakkerSchomerus,Gabelli}. Using this result, we can calculate the NRF, which clearly goes below 1, as represented on Fig. \ref{fig_NRF} where we observe a minimum value of 0.71. This proves the existence of two-mode vacuum amplitude squeezing in the noise emitted by the junction, a direct consequence of the presence of photon pairs. In terms of the current-current correlator, our result is another example of violation by electronic quantum noise of a Cauchy-Schwartz inequality\cite{BBRB}.

\emph{Methods}

$G\inp{f}$, $S_a\inp{f}$, the electron temperature $T$ and the attenuation of the excitation line were all ascertained by measuring and fitting the photo-assisted noise at relatively high ac excitations. This yields $T=20\unit{mK}$ and a setup noise temperature of $5\unit{K}$. Since the signal measured here is 1000 times weaker than that measured in \cite{C4classique}, the experiment was first performed by replacing the sample with a macroscopic $50\unit{\Omega}$ resistor heated with a dc current. For such a sample one expects $G_2=0$. The observed signal thus allowed us to calibrate out the spurious signal of the detection, mainly due to the cross-talk between the digitizer channels.

\emph{Acknoledgements}

We acknowledge fruitful discussions with A. Bednorz, W. Belzig, A. Blais, J. Gabelli, G. Gasse, P. Grangier and F. Qassemi. We thank L. Spietz for providing us with the tunnel junction and G. Lalibert\'e for technical help. This work was supported by the Canada Excellence Research Chairs program, the NSERC, the MDEIE, the FRQNT, the INTRIQ and the Canada Foundation for Innovation.


\begin{thebibliography}{10}
\expandafter\ifx\csname url\endcsname\relax
  \def\url#1{\texttt{#1}}\fi
\expandafter\ifx\csname urlprefix\endcsname\relax\def\urlprefix{URL }\fi
\providecommand{\bibinfo}[2]{#2}
\providecommand{\eprint}[2][]{\url{#2}}

\bibitem{PAN_squeezing}
\bibinfo{author}{Gasse, G.}, \bibinfo{author}{Lupien, C.} \&
  \bibinfo{author}{Reulet, B.}
\newblock \bibinfo{title}{Observation of squeezing in the electron quantum shot
  noise of a tunnel junction}.
\newblock \emph{\bibinfo{journal}{Phys. Rev. Lett.}}
  \textbf{\bibinfo{volume}{111}}, \bibinfo{pages}{136601}
  (\bibinfo{year}{2013}).

\bibitem{Flurin}
\bibinfo{author}{Flurin, E.}, \bibinfo{author}{Roch, N.},
  \bibinfo{author}{Mallet, F.}, \bibinfo{author}{Devoret, M.~H.} \&
  \bibinfo{author}{Huard, B.}
\newblock \bibinfo{title}{Generating entangled microwave radiation over two
  transmission lines}.
\newblock \emph{\bibinfo{journal}{Phys. Rev. Lett.}}
  \textbf{\bibinfo{volume}{109}}, \bibinfo{pages}{183901}
  (\bibinfo{year}{2012}).

\bibitem{Glauber}
\bibinfo{author}{Glauber, R.~J.}
\newblock \bibinfo{title}{Some notes on multiple-boson processes}.
\newblock \emph{\bibinfo{journal}{Phys. Rev.}} \textbf{\bibinfo{volume}{84}},
  \bibinfo{pages}{395--400} (\bibinfo{year}{1951}).

\bibitem{Loudon}
\bibinfo{author}{Loudon, R.}
\newblock \emph{\bibinfo{title}{The quantum theory of light}}
  (\bibinfo{publisher}{Oxford university press, Oxford}, \bibinfo{year}{2000}),
  \bibinfo{edition}{3rd} edn.

\bibitem{C4classique}
\bibinfo{author}{Forgues, J.-C.} \emph{et~al.}
\newblock \bibinfo{title}{Noise intensity-intensity correlations and the fourth
  cumulant of photo-assisted shot noise}.
\newblock \emph{\bibinfo{journal}{Sci. Rep.}} \textbf{\bibinfo{volume}{3}},
  \bibinfo{pages}{2869} (\bibinfo{year}{2013}).

\bibitem{Lesovik}
\bibinfo{author}{Lesovik, G.~B.} \& \bibinfo{author}{Levitov, L.~S.}
\newblock \bibinfo{title}{Noise in an ac biased junction: Nonstationary
  aharonov-bohm effect}.
\newblock \emph{\bibinfo{journal}{Phys. Rev. Lett.}}
  \textbf{\bibinfo{volume}{72}}, \bibinfo{pages}{538--541}
  (\bibinfo{year}{1994}).

\bibitem{GR1}
\bibinfo{author}{Gabelli, J.} \& \bibinfo{author}{Reulet, B.}
\newblock \bibinfo{title}{Dynamics of quantum noise in a tunnel junction under
  ac excitation}.
\newblock \emph{\bibinfo{journal}{Phys. Rev. Lett.}}
  \textbf{\bibinfo{volume}{100}}, \bibinfo{pages}{026601}
  (\bibinfo{year}{2008}).

\bibitem{GR2}
\bibinfo{author}{Gabelli, J.} \& \bibinfo{author}{Reulet, B.}
\newblock \bibinfo{title}{The noise susceptibility of a photo-excited coherent
  conductor.} (\bibinfo{year}{2008}).
\newblock \bibinfo{note}{ArXiv:0801.1432}.

\bibitem{GR3}
\emph{\bibinfo{title}{The noise susceptibility of a coherent conductor}}, vol.
  \bibinfo{volume}{6600} of \emph{\bibinfo{series}{The Proccedings Of SPIE}}.

\bibitem{S3BR}
\bibinfo{author}{Reulet, B.}, \bibinfo{author}{Senzier, J.} \&
  \bibinfo{author}{Prober, D.~E.}
\newblock \bibinfo{title}{Environmental effects in the third moment of voltage
  fluctuations in a tunnel junction}.
\newblock \emph{\bibinfo{journal}{Phys. Rev. Lett.}}
  \textbf{\bibinfo{volume}{91}}, \bibinfo{pages}{196601}
  (\bibinfo{year}{2003}).

\bibitem{Beenakker}
\bibinfo{author}{Beenakker, C. W.~J.}, \bibinfo{author}{Kindermann, M.} \&
  \bibinfo{author}{Nazarov, Y.~V.}
\newblock \bibinfo{title}{Temperature-dependent third cumulant of tunneling
  noise}.
\newblock \emph{\bibinfo{journal}{Phys. Rev. Lett.}}
  \textbf{\bibinfo{volume}{90}}, \bibinfo{pages}{176802}
  (\bibinfo{year}{2003}).

\bibitem{Kindermann}
\bibinfo{author}{Kindermann, M.}, \bibinfo{author}{Nazarov, Y.~V.} \&
  \bibinfo{author}{Beenakker, C. W.~J.}
\newblock \bibinfo{title}{Feedback of the electromagnetic environment on
  current and voltage fluctuations out of equilibrium}.
\newblock \emph{\bibinfo{journal}{Phys. Rev. B}} \textbf{\bibinfo{volume}{69}},
  \bibinfo{pages}{035336} (\bibinfo{year}{2004}).

\bibitem{LesovikLoosen}
\bibinfo{author}{Lesovik, G.} \& \bibinfo{author}{Loosen, R.}
\newblock \bibinfo{title}{On the detection of finite-frequency current
  fluctuations}.
\newblock \emph{\bibinfo{journal}{Journal of Experimental and Theoretical
  Physics Letters}} \textbf{\bibinfo{volume}{65}}, \bibinfo{pages}{295--299}
  (\bibinfo{year}{1997}).

\bibitem{Gavish}
\bibinfo{author}{Gavish, U.}, \bibinfo{author}{Levinson, Y.} \&
  \bibinfo{author}{Imry, Y.}
\newblock \bibinfo{title}{Detection of quantum noise}.
\newblock \emph{\bibinfo{journal}{Phys. Rev. B}} \textbf{\bibinfo{volume}{62}},
  \bibinfo{pages}{R10637--R10640} (\bibinfo{year}{2000}).

\bibitem{Aguado}
\bibinfo{author}{Aguado, R.} \& \bibinfo{author}{Kouwenhoven, L.~P.}
\newblock \bibinfo{title}{Double quantum dots as detectors of high-frequency
  quantum noise in mesoscopic conductors}.
\newblock \emph{\bibinfo{journal}{Phys. Rev. Lett.}}
  \textbf{\bibinfo{volume}{84}}, \bibinfo{pages}{1986--1989}
  (\bibinfo{year}{2000}).

\bibitem{Iskhakov}
\bibinfo{author}{Iskhakov, T.}, \bibinfo{author}{Lopaeva, E.},
  \bibinfo{author}{Penin, A.}, \bibinfo{author}{Rytikov, G.} \&
  \bibinfo{author}{Chekhova, M.}
\newblock \bibinfo{title}{Two methods for detecting nonclassical correlations
  in parametric scattering of light}.
\newblock \emph{\bibinfo{journal}{JETP Letters}} \textbf{\bibinfo{volume}{88}},
  \bibinfo{pages}{660--664} (\bibinfo{year}{2008}).

\bibitem{Agafonov}
\bibinfo{author}{Agafonov, I.~N.}, \bibinfo{author}{Chekhova, M.~V.} \&
  \bibinfo{author}{Leuchs, G.}
\newblock \bibinfo{title}{Two-color bright squeezed vacuum}.
\newblock \emph{\bibinfo{journal}{Phys. Rev. A}} \textbf{\bibinfo{volume}{82}},
  \bibinfo{pages}{011801} (\bibinfo{year}{2010}).

\bibitem{Eichler}
\bibinfo{author}{Eichler, C.} \emph{et~al.}
\newblock \bibinfo{title}{Observation of two-mode squeezing in the microwave
  frequency domain}.
\newblock \emph{\bibinfo{journal}{Phys. Rev. Lett.}}
  \textbf{\bibinfo{volume}{107}}, \bibinfo{pages}{113601}
  (\bibinfo{year}{2011}).

\bibitem{Nguyen}
\bibinfo{author}{Nguyen, F. m.~c.}, \bibinfo{author}{Zakka-Bajjani, E.},
  \bibinfo{author}{Simmonds, R.~W.} \& \bibinfo{author}{Aumentado, J.}
\newblock \bibinfo{title}{Quantum interference between two single photons of
  different microwave frequencies}.
\newblock \emph{\bibinfo{journal}{Phys. Rev. Lett.}}
  \textbf{\bibinfo{volume}{108}}, \bibinfo{pages}{163602}
  (\bibinfo{year}{2012}).

\bibitem{Heidmann}
\bibinfo{author}{Heidmann, A.} \emph{et~al.}
\newblock \bibinfo{title}{Observation of quantum noise reduction on twin laser
  beams}.
\newblock \emph{\bibinfo{journal}{Phys. Rev. Lett.}}
  \textbf{\bibinfo{volume}{59}}, \bibinfo{pages}{2555--2557}
  (\bibinfo{year}{1987}).

\bibitem{Debuisschert}
\bibinfo{author}{Debuisschert, T.}, \bibinfo{author}{Reynaud, S.},
  \bibinfo{author}{Heidmann, A.}, \bibinfo{author}{Giacobino, E.} \&
  \bibinfo{author}{Fabre, C.}
\newblock \bibinfo{title}{Observation of large quantum noise reduction using an
  optical parametric oscillator}.
\newblock \emph{\bibinfo{journal}{Quantum Optics: Journal of the European
  Optical Society Part B}} \textbf{\bibinfo{volume}{1}}, \bibinfo{pages}{3}
  (\bibinfo{year}{1989}).

\bibitem{BeenakkerSchomerus}
\bibinfo{author}{Beenakker, C. W.~J.} \& \bibinfo{author}{Schomerus, H.}
\newblock \bibinfo{title}{Counting statistics of photons produced by electronic
  shot noise}.
\newblock \emph{\bibinfo{journal}{Phys. Rev. Lett.}}
  \textbf{\bibinfo{volume}{86}}, \bibinfo{pages}{700--703}
  (\bibinfo{year}{2001}).

\bibitem{Gabelli}
\bibinfo{author}{Gabelli, J.} \emph{et~al.}
\newblock \bibinfo{title}{Hanbury brown״wiss correlations to probe the
  population statistics of ghz photons emitted by conductors}.
\newblock \emph{\bibinfo{journal}{Phys. Rev. Lett.}}
  \textbf{\bibinfo{volume}{93}}, \bibinfo{pages}{056801}
  (\bibinfo{year}{2004}).

\bibitem{BBRB}
\bibinfo{author}{Bednorz, A.}, \bibinfo{author}{Bruder, C.},
  \bibinfo{author}{Reulet, B.} \& \bibinfo{author}{Belzig, W.}
\newblock \bibinfo{title}{Nonsymmetrized correlations in quantum noninvasive
  measurements}.
\newblock \emph{\bibinfo{journal}{Phys. Rev. Lett.}}
  \textbf{\bibinfo{volume}{110}}, \bibinfo{pages}{250404}
  (\bibinfo{year}{2013}).

\end{thebibliography}
\end{document}